# Consenting to Internet of Things Across Different Social Settings[1]

Yasasvi Hari, Rohit Singh, Kizito Nyuytiymbiy, David Butera
Carnegie Mellon University (CMU)

## Abstract

Devices connected to the Internet of Things (IoT) are rapidly becoming ubiquitous across modern homes, workplaces, and other social environments. While these devices provide users with extensive functionality, they pose significant privacy concerns due to difficulties in consenting to these devices. In this work, we present the results of a pilot study that shows how users consent to devices in common locations in a friend's house in which the user is a guest attending a party. We use this pilot study to indicate a direction for a larger study, which will capture a more granular understanding of how users will consent to a variety of devices placed in different social settings (i.e. a party house owned by a friend, an office space for the user and some 40 other employees, the bathroom of a department store, etc.). Our final contribution of this work will be to build a probability distribution which will indicate how probable a given user is to consent to a device given what sensors it has, where it is, and the awareness and preferences of each user.

## I. Introduction

Today's homeowners are surrounding themselves with smart devices that can make their life easier. For example, Google Home, Amazon Alexa, smart lights, smart plugs, smart TVs, cameras, and many other IoT home appliances, as shown in Fig. 1, permeate the landscape of the modern house. These smart devices, which are used on a daily basis and have the capability to send or receive data through the internet, are called the *Internet of Things (IoT)* [1], as shown in Fig. 1. These IoT devices can be equipped with multiple sensors such as microphones, cameras, gyroscope, infrared scanners, and biometric scanners. These sensors can accomplish a range of tasks, such as e-commerce, surveillance, motion detection, house chores, and even entertainment. With the advent of cheaper IoT devices and improved access to high-speed internet, these devices will become an integral part of our everyday life.

It is envisioned, in the future, these IoT devices will be cheap enough to be densely deployed and interconnected. In a hypothetical scenario, a user will be able to check for grocery items at home and order it online just through their smart watch. Moreover, the user can also automate the process, where the refrigerator or other smart devices can predict, and pre-order items based on the user preferences. These systems will be able to connect people, processes, data, and devices in an intelligent manner leading to a better user experience, which is called Internet of Everything (IoE) [2].

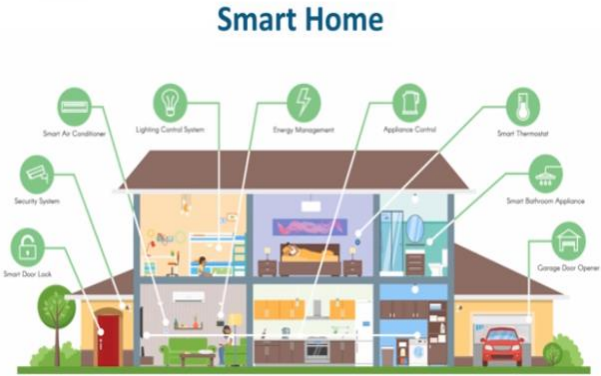

**Fig 1.** Schematic diagram of the traditional home IoT devices used daily [21].

Although IoT (or IoE, in the future) can seem fascinating and crucial for improving user experience, these devices often cause multiple privacy harms [3] to the owner of the device(s) and the individuals surrounding the owner. Theoretically, an owner, to a certain extent, can mitigate these harms by adjusting their privacy settings that suit his/her needs. However, this alteration does not guarantee that harmlessness either inadvertently or advertently to the other people in the room, regardless of who they are. The work we present here as well as the work we aim to do will answer these questions.

### a. Privacy Concerns with IoT

Earlier research shows that people are relatively comfortable with data mining when they are in public spaces. However, even though the home is a private place where users should display increased concerns about their privacy, it would seem that the opposite is true. In this section, we discuss current models of decision making about consent and how these models apply here.

Users often interact with data-hungry IoT devices without fully understanding the full impact of their interactions. Multiple factors make this process challenging for an individual. Our "bounded rationality" [4] limits us from taking rational privacy-sensitive decisions. Humans are not like machines and cannot make binary decisions based on a fixed logic. The logic for a decision changes with time, location, mood, and surrounding individuals. In particular, an individual's decision making is altered based on how much an individual knows. If an individual is fully informed about what data is being shared, who will have access to the data, and how are they using it, then an individual can make a rational decision.

---

[1] This paper was done as the completion for the class project and was funded by CyLab Security & Privacy Institute at CMU. We would like to thank Prof. Lujo Bauer, Prof. Timothy Libert and Prof. Douglas Sicker for their recommendations and suggestions.

Nevertheless, an individual's access to complete information cannot guarantee rational decision making. An enormous amount of data and its inherent complexity can often overwhelm an individual's decision making. Although an individual had complete information and had a perfect understanding of the data, the individual may still make an irrational decision based on the short-term incentive/motivation it is offered in exchange for its decision [5].

It has been shown that there exist multiple privacy concerns in IoT based applications [6], which depends on the type of the device, type of data being shared, and the retention period of the data. It is expected that each of these IoT devices can potentially be interconnected to each other, creating a complicated mesh of data sharing approach. In this process, a lot of the users' data can be collected and analyzed without the direct knowledge of the user. With ultra-dense networks and the Internet of Nano-Things [7] coming in the future, these problems will only be exacerbated. This data can range from information, such as device-identifiable information (DII), personally identifiable information (PII), sensitive personal information, and behavioral information, resulting in multiple individual privacy concerns [8].

Data collection and processing by sensors/IoT devices indoors is perceived differently by different users in different scenarios. For example, users are willing to give up their most sensitive information for a service of their choice, which they would not have done in a normal scenario [5]. On the other hand, it is challenging to collect audio and video data in classrooms, without causing any privacy harms to improve teaching/education systems [9]. One solution is to train using privacy behavior [10] or improve device design and best practices (i.e., privacy by design) [11]. The other solution is to allow the owners of these devices to adjust their privacy settings [12].

Privacy harms can still exist for devices installed outside an owners' home. Some devices are hidden from plain sight, so it can be hard to understand their data collection. Although an individual can alter the privacy settings of their own devices, it is hard or nearly impossible for an individual to adjust the privacy settings of the IoT devices installed at schools, stores, offices, and at a friend's home. An individual who does not own that IoT device and to prevent further privacy harms the individual can either (a) request the owner to change the privacy settings to that individual's liking, or (b) the individual has to leave that room or area (we term this as a barrier to entry), or (c) the individual changes his/her decision and adjusts with the surrounding context. The first option is generally avoided by an individual since it can either lead to embarrassment for an individual in a social setting, or the decision is at the disposal of the owner, who is unlikely to change the privacy settings. The latter options are the most likely solutions practiced by individuals aware of these IoT devices and these harms.

### b. Our Contribution

In this paper, we examine the consent of users to allow IoT devices to collect and process their personal data. Our discussion does not consider how the participants consent to devices they own in their own home, since if they decide to not consent, then they can remove the device or disable it. We envision their preferences to consent will change based on who, when, and how, the data is collected. Specifically, we will answer the following questions:

1. What factors motivate users to not consent to IoT devices in one-on-one scenarios and do these factors change in group scenarios?
2. Do IoT devices ever pose a barrier to entry for a given user?
3. How do users negotiate to disable IoT devices which they feel to be intrusive or unwanted?

If we had more time and resources to complete a more robust survey, we would have more comprehensive answers to these questions. However, since we did not, we focused primarily on gaining insights onto general insights on specific situations which represent a partial answer to the first question. We discuss this further in sections 3, 4, and 7.

Questions 1 and 2 can be further broken down in three ways to gain more granular insights about the attitudes to consenting to IoT devices: sensor type, location, and number of other people. Specifically, we define a classification system based on the type of sensors each device uses. From this classification, we build an understanding of users' preferences to these types of sensors. This classification also allows us to generalize these privacy preferences to future devices which use these types of sensors.

We build upon existing work done by Emami-Naeini et. al [12]. We further try to understand how the location of these sensors impacts the decision of a user to consent to an IoT device. For example, the same work provides a table of the percentages of users who consent to an IoT device in their house. Our work expands this idea of location by asking users where in the house they would feel that a given device is most uncomfortable (i.e. in the bathroom, in the hallway, in the bedroom, etc.). As such, our contribution makes this understanding more granular.

Finally, the last way in which we further existing work is by factoring in the number of other people in the setting. For example, does the probability of a user consenting to a particular device in a particular location change if the number of others in the room?

Taking these three considerations in tandem, we aim to produce a probability distribution which returns the probability that a given user will consent to the device given its classification, its location, and how many other people are in the setting. Using this distribution, an end user can better understand where they can place a given device, an interior designer can better plan which devices can fit a specified location, and a device developer can better understand user concerns and barriers to use.

## II. Related Work

IoT devices have scaled into a litany of places in the home, and as such take on a variety of applications. For example, consider the surveillance of nannies by the parents of their children [13]. There are plenty of justifications for keeping a nanny cam, many of which the nannies themselves agree as reasonable. Despite these valid justifications, the question is whether the nannies should have to consent to the camera being there. If they do consent to the camera, do they consent to their audio being recorded as well, as in the case of Angella Foster? If that audio is then used for other applications downstream, must the nannies consent to those applications as well? Does accepting the offer of employment to work as a nanny in the house imply this consent, and does not consenting imply turning down the job?

How are people who enter the house, such as nannies, notified about how these devices and what they do? Emami-Naeini et. al. examine this question in detail and identify specific categories of devices and data collection for which users want to be notified in more detail [12]. Particularly, one relevant finding from their work demonstrates that for applications which collect sensitive data (i.e. biometric data) induce more discomfort in the minds of users than less personal data such as environmental data (i.e. temperature).

At what point concerns constitute a barrier for entry into the house? If they do, how do people like nannies, who rely on being in a given building for employment, continue to find employment if they cannot enter the building? Even if they do not pose such a barrier, are these people reluctant to do their jobs because of these devices?

When we buy an IoT device we can alter the setting which suits our needs and to be in line with our privacy preferences. However, this does not mean that other individuals (family, friends, house help, neighbors or strangers) also share the device owner's privacy settings. Recently, Alexa [14] updated their privacy setting through which users can view, hear, and delete Alexa's voice recordings. Samsung has also experienced a similar issue with their Smart TVs and voice commands, but without the option to disable voice commands [15]. From this, we see another privacy problem.

Suppose that you have two or more IoT systems with different levels of configurability (i.e. a system which you can completely disable the feature deemed to be intrusive and another where you can only partially disable a feature you dislike). Will this configurability of these systems dissuade or persuade people from entering the building? How will such a decision impact the dynamic of the social group?

Such physical systems which collect data have been studied through the lens of user consent and privacy [16-20]. An interesting formalization of policies pertaining to the disclosure of user data collected by sensors is articulated by Mehrotra et. al., who postulates that an "overarching policy" could enable disclosure of information of the party (i.e. the number of people in a given room at a given time) but prevent the disclosure of information relating to specific people (i.e. the identities of people in a given room). Such a privacy policy does seem to be sufficient for most users, and consistent with the findings of Emami-Naeini and her group [12] as it does not expose sensitive information. It would be interesting from our perspective to build a ranking of which kinds of data will create the barriers to entry which we were previously discussing. For example, we could ask participants to imagine going to their friends' house with different sensors which collect biometric information and if that would constitute a barrier to their entry. Then we could ask them to compare this scenario to a similar scenario where instead of biometric data, the sensors in the house would be nanny cams, for example. We would then ask if these nanny cams would be more or less likely to constitute a barrier to entry into the house. By doing this iteratively over the set of possible sensor inputs, we can better understand what it would take to create the barrier for entry into the building. Achieving this would also be an important step in understanding which sensors users would find acceptable and thus help us understand for and against which sensors guests or visitors could potentially negotiate. For example, a nanny might find audio recording to constitute a barrier of entry into a given house but could consent to there being video recording. Should his or her employer agree, then the barrier of entry no longer exists, and everyone's privacy preferences are respected.

Although most of the debate for privacy harms has been centered on "notice and choice", users still suffer from traditional privacy harms. To protect users from these harms, extensive legislation is typically passed. A good example is the privacy implementations on websites that followed the EU's GDPR. Companies were compelled to comply with legislation while also satisfying users of their services. This gave rise to the advent of cookies consent interfaces to obtain user consent before collecting information about them.

The case is quite different with IoT. No legislation exists about consent for IoT-enabled, data-hungry devices. Consequently, there is currently only negligible concern about user consent before allowing IoT devices collect information. Industry trends also show no plans for such consent implementations in the near future. The need for home IoT consent becomes more serious when you consider the fact that unlike the cookies consent where users go to sites for some service of interest to them, visitors to an IoT-enabled home may not want to be remembered by the IoT devices. An example is a connected baby monitor with a camera that captures video and audio (in some cases) to help parents keep an eye on the baby in real time [13]. While this is technology is useful, anyone who comes into the field of the camera's view, consensually or otherwise, will be viewed by the camera.

We conclude our literature by asking the following questions. First, does a user's decision (i.e. the decision of a guest, employee, or passerby) to not consent to the IoT device's data collection create a barrier to entry of that building for them? If it does create such a barrier, how can different users negotiate a configuration of the system such that its data collection is both useful and respectful of everyone's privacy preferences? Particularly, we are concerned with settings of small groups, for example a small party or gathering of 5 or 6 people. We discuss our approach to answer these questions in the next section.

## III. Our Approach

There is a growing research interest to predict users' privacy preferences based on a broad range of attitudes and scenarios [12]. In this work, we analyze granular scenarios to justify the varying user preferences. Our contextual questions flow naturally from the work done in [12] by adding granularity to their questions. We achieve this granularity in four ways as shown in Fig. 2 and listed below:

1. Type of Data: Despite the fact that the market for IoT devices is diverse, we can classify these devices based on their sensors. Using this classification, we can ensure that our work is more generalizable and not device-specific. For example, we ask questions about Amazon Alexa in each of the contextual questions, but we map Alexa to its type of sensor in our analysis, namely 'microphone'. While Alexa might be a popular device now, there is no guarantee that it will be as popular in the future, which is why we want to map it this way.

2. Type of Owner: In some cases, even if the data is not too sensitive, the privacy preference of a user might change based on the owner or who collects this data. For example, imagine a party scenario where a given user runs into a former romantic partner with whom the relationship badly. This scenario can be awkward, and so the user may not consent to this device.

3. Type of Scenarios: The privacy preferences also change massively in the case of varying social settings in which these device interactions occur. Adding more specific sublocations (e.g., the bedroom, the restroom, and the hallway) to these social settings, might further complicate these preferences.

4. Type of Retention Period: Users are often fine with devices using their personal data for a small duration of time, such as financial data for buying goods and services. However, if the data is retained for a longer period, users' preferences might become more conservative even for least sensitive data.

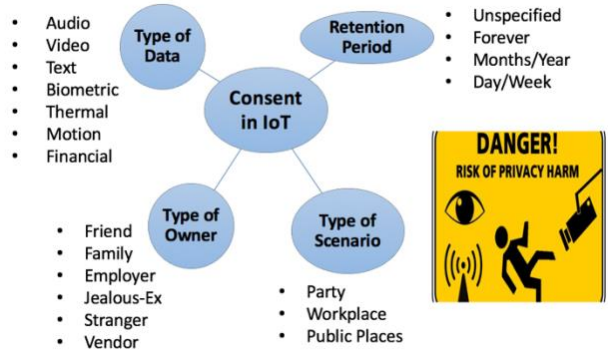

**Fig 2.** Types of factors affecting users' consent to IoT devices to avoid the risk of privacy harm.

These 4 classes and sub-classes can result in more than 400 different combinations where a user might feel the risk of privacy harm. Due to our lack of funding and time we only explore one such combination (i.e., type of data: audio, type of owner: friend, type of scenario: party, and retention period: unspecified). In particular, we focused on the setting of a friend's house. We term this as our *location* and the *sublocations* we surveyed are the hallway, restroom, and bedroom of the friend's house. The need for this additional granularity comes from the fact that certain users may feel comfortable in one setting but not in others. This is best captured by the nanny cam scenario [13].

## IV. Methodology

To collect our data, we ran an Amazon Mechanical Turk (Mturk) Human Intelligence Task (HIT) and used Google Forms for our pilot survey. For the sake of brevity and reducing the cognitive load on our participants, we kept our survey limited to 12 total questions, 2 of which were only asked if the previous questions were answered in a specific way. We user tested the survey on 25 participants, and they took on average approximately 5-7 minutes to complete the survey. To give our participants in the Mturk survey more time, we assumed that it would take approximately 10 minutes to complete. From this estimate, we used the current U.S. minimum wage of $7.25 and rounded up to pay out $2 to each participant after the completion of the survey. For the final survey conducted on Mturk, we got 73 total respondents. We attach the survey in the appendices section.

### a. Survey Structure

We started the survey with a standard consent form, as seen in Appendix 1. After this, we asked if the participants owned IoT devices in their own homes and if so, which ones. The next question asked if anyone of their friends owned IoT devices, and if so, which ones. These two questions help us understand how aware the participant is of our data.

Once we gauged the participant's awareness, we proceeded to try to understand their concerns with these devices. Question 3 allowed respondents to give several reasons why they believe their privacy might be violated from a list of potential reasons. It also gave them a text box to indicate a reason that was not listed. This question builds on the previous question by now gauging how aware the respondents are of how some standard privacy concerns, and also giving them a place to voice their own concerns about these devices. Questions 4 through 6 followed the structure outlined in Section 3. Immediately after this question, we followed by asking how they would have liked to have been notified about the device, again providing a text box to clarify other notification methods which they would like to see. This question gives us some insight into how future devices can be developed to avoid these concerns about consent notifications.

Question 8 is the final question before the demographic questions and deserves some special attention. The question asks users to consider a situation where the user finds out that Alexa recorded the entire conversation a few days after the party. The question asks the respondents if they would like the transcript to be deleted, without any indication of what the transcript actually says. We wanted to use this question to see if the user fears saying or doing anything which they may regret, and specifically if they fear the device recording them. This question achieves this goal without biasing the respondent's answer since it does not place indicate what the user might have said or did that they later would regret.

### b. Demographics and Respondents

For our survey, we recruited participants on MTurk who were both greater than the age of 18 and registered as Amazon Masters, or respondents who consistently pass Amazon's statistical measures of quality.

### c. Quality Control

Our primary goal in terms of quality control was to reduce the amount of cognitive load on each participant. For this reason, we limited our survey to 3 scenarios, each with an easily identifiable device (e.g. Amazon Alexa). We also achieved a reduced cognitive load by frontloading the more taxing questions about the user's consent preferences and placing the questions about demographics at the end. By structuring the survey in this manner, we decrease the likelihood that the respondent gets tired or frustrated and submits a nonsensical answer. Since our survey was so short, we did not feel the need to add attention checks. When we run the longer, more comprehensive survey, this will be necessary.

## V. Analysis

Below, we analyze the data we collected over our Mturk HIT. The preliminary results indicate interesting user preferences and concerns about users and their interaction with IoT devices. The results shown in this section will work as a motivation for our future studies.

### a) Privacy Preference Question Analysis

Before we move on to consent analysis, we need to understand the ownership and privacy preferences of users, which has been collected using Questions 1 through 6, shown in Appendix A. Figure 3 shows the number of participants: (a) who own IoT devices, i.e., "Owner", and (b) who are friends with people who own IoT devices, i.e., "Friends". and who know friends who "maybe" own IoT devices.

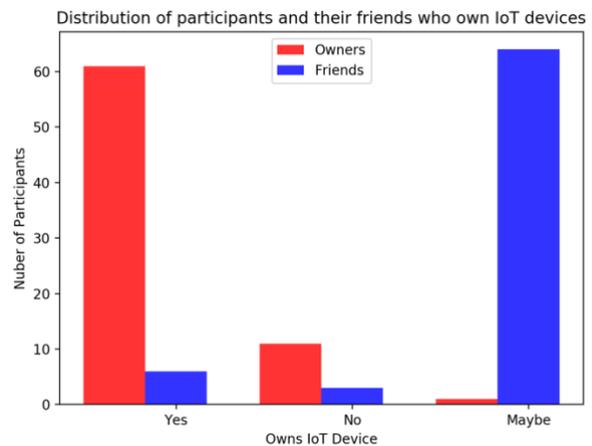

**Fig 3.** Understanding the ownership of IoT devices among participants and their friends.

As noted in Section 4, these responses illustrate the awareness of the user to IoT devices. Nearly 60 users said that they themselves owned IoT device(s), which speaks to the prevalence of these devices as well as the willingness of users to keep these devices in their homes or with them in general. Further we also ask participants if they are friends with people who own IoT devices. This question serves two purposes: (a) get a sense if participants of aware of presence of IoT devices outside their home, and (b) it works as a foundation for the later questions in the survey. It seems that most of the participants are unaware if their friends own any IoT devices. Table 1 further classifies the type of devices owned by the participants and their friends.

Interestingly the response from Fig.3 about friends owning IoT devices does not match with the results shown in Table 3. This can be due to the lack of understanding the exact definition of IoT or the presence of lazy and uninterested users. The latter is unlikely because a lazy user would have to select more options in Question 4.

**Table 1:** List of IoT Devices Owned

| Devices | Number Owned by Respondents | Number of Devices Owned by Respondents' Friends |
|---|---|---|
| Smart TV | 49 | 56 |
| Personal Home Assistant | 40 | 56 |
| Fitness Tracker | 27 | 48 |
| Smart Thermostat | 17 | 29 |
| Smart Doorbell | 14 | 30 |
| VR Headsets | 10 | 18 |
| Smart Switch/Light | 4 | 0 |
| Pet Tracker | 2 | 8 |
| Others | 3 | 12 |
| None | 3 | 1 |

Now that we have established the prevalence of these IoT devices in the lives of ourselves and our friends, we turn to the concerns that the users may have about their privacy. From Figure 4 above, we highlight the two most interesting and common fears. Almost 63% of respondents indicated that they are concerned that the data these devices collect may "end up in the wrong hands" (i.e., Type 3), and 57.5% of respondents indicate that they worry about excessive data collection (i.e., Type 1). Together, these responses indicate a general fear in the minds of those surveyed about a ubiquitous set of applications which do not have a clear remit of what types of data they are allowed to collect.

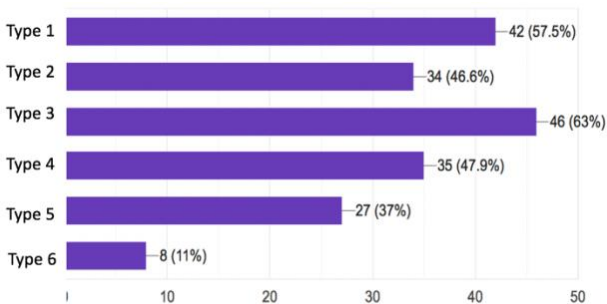

**Fig 4.** Different types of privacy concerns among participants. Type 1:``I feel that they collect too much information/data than needed", Type 2:``I don't have control over the data they collect", Type 3:`` I fear the data might end up in the wrong hands", Type 4:``I don't have control over who has access to the data", Type 5:`` The vendor or manufacturer is using the data in ways I did not expect", and Type 6:``None"

Interestingly the results of Figures 3 and 4 seem to be contradictory. If the users fear some nebulous class of applications and their supposedly sinister data collection and aggregation, why buy these applications in the first place? Why buy precisely the thing which we fear and place it in our own homes, and in some cases, even on our bodies? It seems that the users' fears and actions are at odds here.

One possible explanation (which will be a question in a future, more complete survey) would be that the users feel that the benefits gained in the functionality of these devices outweigh the privacy harms caused by these devices. Another approach could be to ask respondents to rank which device's benefits most outweigh their privacy harms and why. These questions not only help us to understand this contradiction but also can help us understand how users may or may not consent to IoT devices which they do not own. For example, if a given user perceives a device to be "useful", then they might be more willing to consent to the device despite knowing the potential harms that they may incur.

Another potential reason for this contradiction is that the users may not feel that they have any real control over their consent in the first place. If everyone around a given user has some such device connected to some network of other devices, then that the user may believe that their consent means little in the grander context of the network. If I am surrounded by these IoT devices, and each of them collects some unknowable amount of data and transmits it throughout its own network discreetly, how much power do I really have to stop them? The next survey we conduct will also ask if users feel empowered to make decisions about their own consent with regards to these IoT devices.

Interestingly, the first explanation may stem from a misevaluation of how costly privacy harms may actually be. As Solove notes [3], people may have difficulty quantifying a privacy harm, which may be skewing their cost-benefit analysis of these devices. If a user perceives significant physical harm stemming from some sort of device, they are probably unlikely to consent to it. Privacy harms are distinct from physical harms because often times, they are much more subtle and harder to fully understand. If you notify a user that their data is being collected, it seems to only breed more questions. What type of data is being collected? Where is the device that is collecting it? How long will this data be stored? With whom will it be shared? How am I harmed from this? Even if we have all of these answers, how do we reason about the tradeoff between functionality and privacy?

One way to partially solve some these problems is to notify the users meaningfully. If the device owner can answer a subset of questions (e.g. what type of data is being collected, what the data retention time is, etc.), other users could potentially make more informed consent decisions. We polled the respondents about how they preferred to be notified of the device in Figure 5.

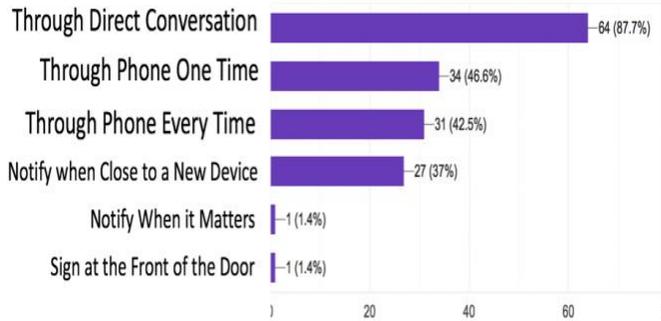

**Fig 5.** Ways through which a participant can be notified about an IoT device

The respondents seem to place a lot of trust on the device owner. The most frequent selection was by far through a direct conversation with the device owner, but this could be inconsistent if the owner does not inform each user in the same way. Also, the owner might be unclear about the specific configurations about their device and how these configurations can impact privacy, and so there could be some misinformation which could negatively impact the consent decision.

One interesting suggestion is notification through cell phones. This way, we can actually standardize the types of notices that we can give all users. Our next study should capture what types of questions they have about these devices and how a notification could be most useful.

We turn now to our analysis of our contextual concerns.

b) **Contextual Scenario Analysis**
The primary goal of each of these contextual questions was to collect context-dependent data about how we expect the users to behave in these scenarios.

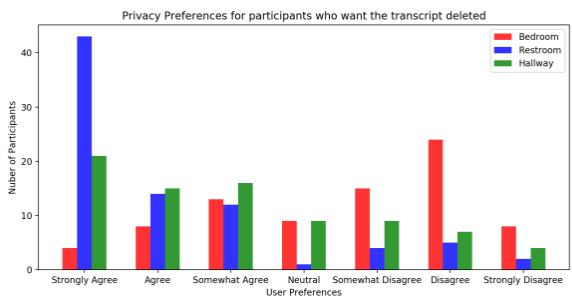

**Fig 6.** The above chart shows the distribution of responses of users who said that they want the transcript deleted.

From this preliminary analysis, we can see immediately that the perceived risk is dramatically different across different sub-locations. This indicates that our initial assumption that consent to IoT devices is dependent on the sub-location. Despite being in a familiar location (their friend's house), 80.8% of respondents said that they wanted the transcript compiled by Alexa deleted. Fig. 6 shows the responses of such respondents to the contextual questions. Clearly, these users have significant privacy concerns (as evidenced by the fact that they want the transcript deleted), which they seem to have felt most acutely in the bathroom. This finding concurs with previous work done in the same field [12].

## VII. Limitations

Our pilot study was not meant to be a conclusive study; we simply do not have the data to exhaust all of our scenarios. We do not expect this data to be sufficient to draw any real conclusions. Rather, our goal was to begin our exploration into this topic.

Prior work has noted some potential problems with Amazon Mturk, namely that it could sample from a biased population. At the same time, other work has also noted that the results from Mturk are similar to other online sources, which seems to suggest that perhaps the data are not biased.

More generally, the limitations of a survey stem from its abstraction. We are asking the respondents to *imagine* scenarios where they interact with these devices. Whether or not they actually behave in the same way in real life still remains to be seen. We believe that one way to remedy this limitation is to run a focus group study in which participants role play the scenarios described in the survey and consider how they feel after. Role playing should give us a closer approximation of real-life behavior.

Regarding our pilot survey, we are worried about the ordering effect. We did not randomize the order that the participants saw these questions, and so we cannot estimate the impact that a given user's attitude towards a given contextual scenario question has on the ones that came after it.

## VI. Conclusion

We conclude by reiterating the three questions which we aim to answer:
1. What factors motivate users to not consent to IoT devices in one-on-one scenarios and do these factors change in group scenarios?
2. Do IoT devices ever pose a barrier to entry for a given user?
3. How do users negotiate to disable IoT devices which they feel to be intrusive or unwanted?

Despite the cost-prohibitive aspects of this study, we believe that this survey points at a larger, more interesting avenue of work which will answer these questions. The probability distribution which we will build creates more robust predictions of how laypeople will and will not consent to IoT devices across common locations in different group settings. This distribution gives quantitative answers to the first two questions. Further, the data collected from the focus groups will nuance the data by providing qualitative data. This data will better capture how people negotiate how these devices impact their privacy. The result of the focus groups will provide a different approach to answering the second question as well as completely answering the third question.

Taken in tandem, we believe that these answers provide a novel, interesting, and significant to the fields of privacy, data collection, and IoT. Future work would build on this contribution by considering more scenarios, different types of devices, and different social settings. We envision that a later stage we will be able to answer more detailed questions regarding consent in IoT appliances, such as:
a) How should an opt-out interface look like?
b) Should there be a consent transfer among devices? In other words, does consenting to one IoT device imply consent to any other device that consumes data from the first device?
c) Which privacy features[2] should the owners be able to control?
d) What privacy features should the non-owners be able to control?
e) Which privacy features should be default?
f) Who should design this choice mechanism: industry or regulators?

## Appendix A

Following are the list of survey questions which we deployed on MTurk

Q1) Do you own any IoT devices at your home?
Options: Yes/No

Q2) Which IoT product(s) do you own, if any? Select all that apply.
- Personal fitness tracker (i.e. Fitbit)
- Smart refrigerator
- Home security monitoring system
- Smart thermostat
- Amazon Echo or similar
- Pet tracker
- Virtual reality headset
- Smart doorbell
- Other (please specify)<Text Box>
- None

Q3) Do any of your friends own any IoT devices?
Options: Yes/No

Q4) Which IoT product(s) does you friend(s) own, if any? Select all that apply.
Options: Same as Q2

Q5) Do you know any privacy harm that these IoT devices can cause? <Text Box>

Q6) Which of these concerns do you relate with? Select all that apply.
- I feel that they collect too much information/data than needed
- I don't have control over the data they collect
- I fear the data might end up in the wrong hands
- I don't have control over who has access to the data

---

[2] Privacy features are settings which allow users to encode their preferences in the device, answering questions such as what data is collected, when should it be collected, who collects it, and how long it should be retained.

- The vendor or manufacturer is using the data in ways I did not expect
- Other (specify)<Text Box>

Q7) The following questions ask you to imagine that you are at a close friend's house for a party of 7-8 people. Your friend has many IoT device at home and he/she likes to control them through voice command using Amazon Alexa/Google Home. He/she has many of these voice controlled devices distributed at different locations in his/her house. The devices can actively hear and record voice commands.

Q7a) Now imagine that you are at the Restroom and you notice an Amazon Alexa/Google Home. Do you feel there is any privacy issues associated with the presence of this device?
Options: Strongly agree/Agree/Somewhat agree/Neither agree nor disagree/Somewhat disagree/Disagree/Strongly disagree

Q7b) Now imagine that you are at the Hallway and you notice an Amazon Alexa/Google Home. Do you feel there is any privacy issues associated with the presence of this device?
Options: Strongly agree/Agree/Somewhat agree/Neither agree nor disagree/Somewhat disagree/Disagree/Strongly disagree

Q7c) Now imagine that you are at the Bedroom and you notice an Amazon Alexa/Google Home. Do you feel there is any privacy issues associated with the presence of this device?
Options: Strongly agree/Agree/Somewhat agree/Neither agree nor disagree/Somewhat disagree/Disagree/Strongly disagree

Q7d) Would you prefer to be informed about the presence of Amazon Alexa/Google Home before you enter the house? Before you enter the room? At all?
Options: Strongly agree/Agree/Somewhat agree/Neither agree nor disagree/Somewhat disagree/Disagree/Strongly disagree

Q7e) If you had that choice, would you allow or deny this data collection?
Options: Allow/Deny

Q7f) How would you want this notification of consent
- I want my friend to notify me
- I would want my mobile phone to notify me only the first time this data collection occurs.
- I would want my mobile phone to notify me once in a while this data collection occurs.
- I would want my mobile phone to notify me every time such data is collected

Q7e) Do you have any other privacy concerns related to this Scenario? <Text Box>

Q8) Now imagine, a few days later, you come to know that Amazon Alexa/Google Home recorded your voice and has transcripts of your conversation on that day
Q8a) Are you fine with your friend having this transcript?
Options: Yes/No

Q8b) On a scale of 1-7, how much do you agree with the following statement: The transcript should be deleted.
Options: Strongly agree/Agree/Somewhat agree/Neither agree nor disagree/Somewhat disagree/Disagree/Strongly disagree

Q8c) Which of these reflect your reaction to your friend's Amazon Alexa/Google Home recording you? Select all that apply?
- I don't mind if there is Alexa recording me or not.
- I mind being recorded, turn the camera off completely.
- I would like to have the recording after.
- I would like some sections of the party recorded but not all.
- Other <Text Box>

Q9) How old are you? <Text Box>

Q10) Gender <Text Box>

Q11)What is the highest degree you have earned?
 Options: (No high school degree/ High school degree / College degree/ Professional degree (masters/PhD)/Prefer not to answer)

Q12) What is your income range?
- Less than $15,000/ year,
- $15,000/ year - $24,999/year,
- $25,000/ year - $34,999/ year,
- $35,000/ year - $49,999/ year,
- $50,000/ year - $74,999/ year,
- $75,000/ year - $99,999/ year,
- $100,000/ year - $149,999/year,
- $150,000/year- $199,999/ year,
- $200,000/ year and above,
- Prefer not to answer

Q15) Suggestions/Comments <Text Box>